# A very early origin of isotopically distinct nitrogen in inner Solar System protoplanets

Damanveer S. Grewal[1]*, Rajdeep Dasgupta[1], Bernard Marty[2]

[1]Department of Earth, Environmental and Planetary Sciences, Rice University, 6100 Main Street, MS 126, Houston, TX 77005

[2]Université de Lorraine, CNRS, CRPG, F-54000 Nancy, France

*correspondence: dsg10@rice.edu

**Understanding the origin of life-essential volatiles like nitrogen (N) in the Solar System and beyond is critical to evaluate the potential habitability of rocky planets[1–5]. Whether the inner Solar System planets accreted these volatiles from their inception or had an exogenous delivery from the outer Solar System is, however, not well understood. Using previously published data of nucleosynthetic anomalies of Ni, Mo, W and Ru in iron meteorites along with their $^{15}N/^{14}N$ ratios, here we show that the earliest formed protoplanets in the inner and outer protoplanetary disk accreted isotopically distinct N. While the Sun and Jupiter captured N from nebular gas[6], concomitantly growing protoplanets in the inner and outer disk possibly sourced their N from organics and/or dust – with each reservoir having a different N isotopic composition. A distinct N isotopic signature of the inner Solar System protoplanets coupled with their rapid accretion[7,8] suggests that non-nebular, isotopically processed N was ubiquitous in their growth zone at ~0-0.3 Myr after the formation of CAIs. Because $^{15}N/^{14}N$ ratio of the bulk silicate Earth falls between that of inner and outer Solar System reservoirs, we infer that N in the present-day rocky planets represents a mixture of both inner and outer Solar System material.**

The origin of life-essential volatiles in the rocky bodies of the inner Solar System is a subject of major debate. Owing to a broad similarity of $^{15}N/^{14}N$ and D/H ratios in the rocky bodies and carbonaceous chondrites, these volatiles are generally believed to have been delivered primarily via volatile-rich carbonaceous chondrite-like material from the outer Solar System[3,4]. The outer Solar System material was displaced inwards within the first few Myr of Solar System formation either due to the perturbations in the protoplanetary disk caused by the growth[9] or migration[10] of giant planets, or admixing of pristine outer Solar System material associated with mass accretion of the proto-Sun[11]. However, affinity of $^{15}N/^{14}N$ and D/H ratios in Earth's primitive mantle with enstatite chondrites (isotopically similar to the primary building blocks of inner Solar System planets[12]) suggests that some portion of volatiles in the inner Solar System planets may have been sourced from the inner Solar System reservoir[1,13]. Although enstatite chondrites contain N, C, and H in significant amounts[1,14], their rather late accretion (~2 Myr after the formation of CAIs)[15] precludes them from providing information on the presence of volatiles in the rocky planet forming material of the inner disk at the very beginning of the formation of the Solar System. However, the earliest volatile history can be investigated by looking at iron meteorites which represent the cores of the earliest-formed protoplanets (possibly within ~0.3 Myr of the formation of CAIs[7,8]). Because the seeds of present-day rocky planets also accreted in the inner disk at similar timescales, a volatile-free or a volatile-bearing nature of these protoplanets has important implications on the dynamics of planetary growth.

The recent discovery of a bimodal distribution of the nucleosynthetic anomalies of several non-volatile elements on meteorite and planetary scales has revolutionized our understanding of the initial architecture of the Solar System[8,16]. The dichotomy necessitates this separation of the protoplanetary disk into two isotopically distinct reservoirs (carbonaceous: CC and non-carbonaceous: NC)[16]. The separation has been linked to the opening up of a gap in the disk either due to the rapid growth of Jupiter's core[8] or a pressure maximum just beyond its orbit[17]. The identification of CC-NC dichotomy in iron meteorites pins the spatial isolation of CC and NC reservoirs within ~1 Myr of CAI formation[8] because accretion and differentiation of their parent bodies (based on Hf-W chronometry) occurred at even shorter timescales[7]. The protoplanets that formed in the NC reservoir (within Jupiter's orbit) and CC reservoir (beyond Jupiter's orbit) accreted in volatile-poor inner disk and volatile-rich outer disk, respectively[18]. However, no study till date has extended the application of the CC-NC dichotomy to a major volatile element to show whether, akin to non-volatile elements, isotopically distinct volatile reservoirs were present in the inner and outer disk. An evidence for the presence of a volatile bearing reservoir with a distinct isotopic composition in the inner disk has important implications on the dynamics of Solar System evolution within the first tens to hundreds of thousands of years. It can be used to date the origin of volatiles in the accretion zone of protoplanets in the inner disk and subsequently infer the contribution of inner and outer Solar System reservoirs to the volatile budget of present-day rocky planets.

Compared to H (not present in all classes of meteorites), and C, (having limited isotopic variation in meteorites; $\delta^{13}C$ <40 ‰), N is present in all classes of meteorites and displays one of the largest isotopic variations ($\delta^{15}N$ >500 ‰; $\delta^{15}N=[\frac{\frac{15N}{14N}_{sample}}{\frac{15N}{14N}_{atm}} - 1] \times 1000$ where '$\frac{15N}{14N}_{atm}$' refers to the isotopic composition of N in Earth's atmosphere ($3.676 \times 10^{-3}$)) in both undifferentiated and differentiated meteorites[14]. In contrast to chondrites and achondrites, which are disequilibrium aggregates of various phases formed by different processes over varying timescales and have their $^{15}N/^{14}N$ ratios affected by thermal and aqueous alteration[3], iron meteorites preserve primitive $^{15}N/^{14}N$ ratios better as N is hosted in chemically resistant Fe, Ni alloy phases (kamacite and taenite)[19,20]. Therefore, iron meteorites more reliably track the formation of the earliest forming rocky body reservoirs[7]. In Figure 1, Extended Data Figs. 1 and 2, and Supplementary Table 1, we show that the $\delta^{15}N$ of iron meteorites vary between –95 ‰ and +164 ‰. $\delta^{15}N$ of each iron meteorite group (thought to represent a distinct parent body) lie in small clusters, while N abundances vary from 0.12 to 131 ppm with each group showing variable distribution. N isotope composition of iron meteorites and its relation, or lack thereof, with the accretion zones of CC-NC reservoirs provides the best candidate to explore the distribution of volatiles in the protoplanetary disk because of the large inter-group variations of their $^{15}N/^{14}N$ ratios. These variations cannot be explained by any physical or chemical mass-dependent fractionation during planetary processing (see Methods for details), and may reflect either N isotope heterogeneities in the molecular cloud[19,21] or a heritage of local self-shielding processes at the surface of the protoplanetary disk[22].

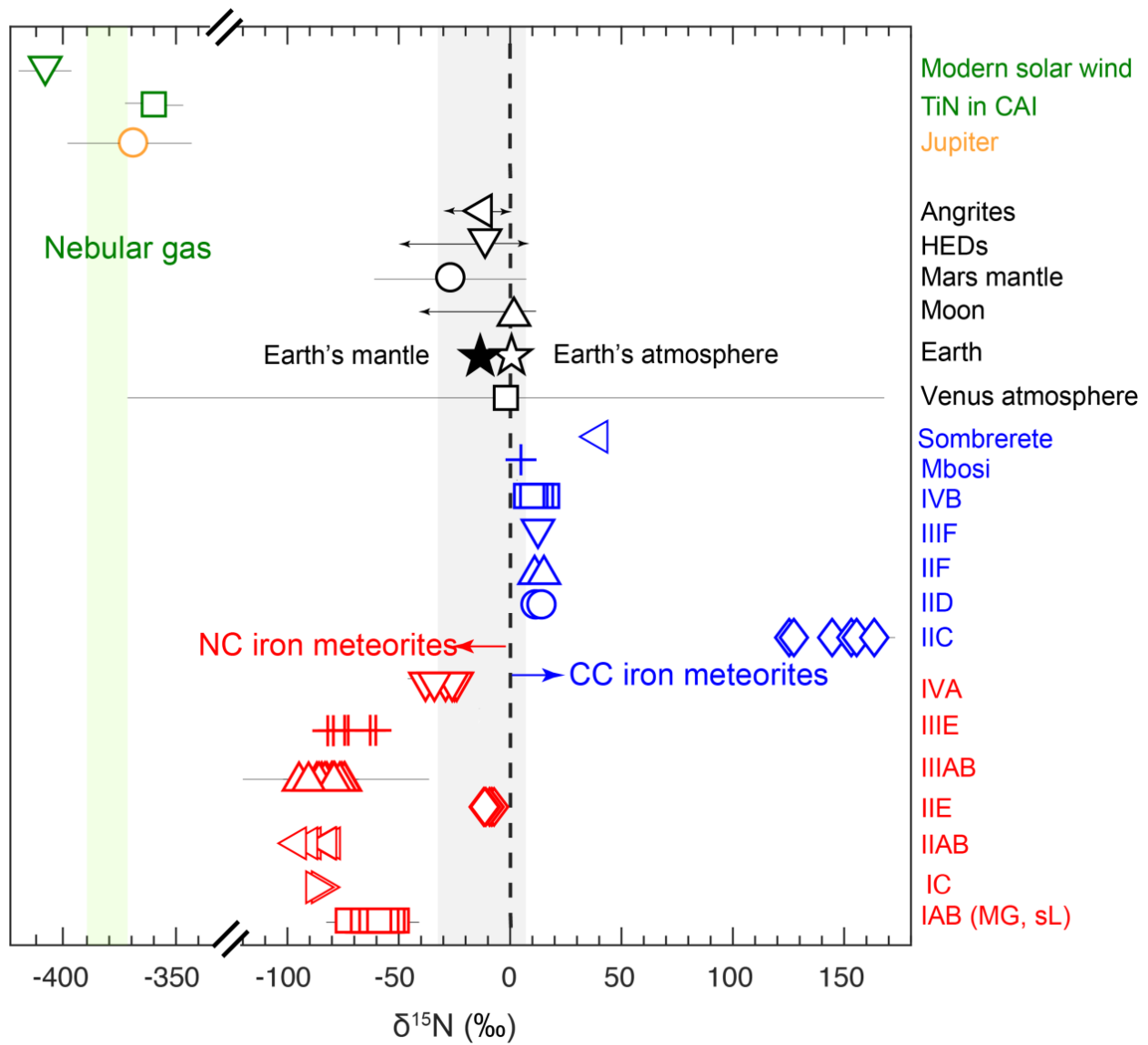

*Figure 1: Variations in $^{15}N/^{14}N$ ratios of various Solar System objects and reservoirs. Iron meteorites and all other rocky body reservoirs are $^{15}N$-rich relative to nebular gas. $\delta^{15}N$ of different groups of iron meteorites are plotted using the CC and NC classification of iron meteorites based on the nucleosynthetic anomalies of Mo isotopes[8,18]. NC iron meteorites have $\delta^{15}N < 0$ ‰, while CC iron meteorites have $\delta^{15}N > 0$ ‰. $\delta^{15}N$ of Earth's atmosphere = 0 ‰. $\delta^{15}N$ of Earth's primitive mantle is estimated to be $-5 \pm 4$ ‰ (2 s.d.)[33]. $\delta^{15}N$ of the bulk silicate reservoirs of all other rocky bodies in the present-day inner Solar System are spread around $\delta^{15}N = 0$ ‰ (marked with a grey band). $\delta^{15}N$ of nebular gas ($-383 \pm 8$ ‰; 2 s.d.) is defined by the measurement of the modern Solar Wind[6]. A similarly $^{15}N$-poor nature of nebular gas is*

*constrained by TiN (the first N-bearing solid phase to condense from nebular gas) hosted in CAI[24]. The atmosphere of Jupiter, presumably sourced from nebular gas, is also $^{15}$N-poor[23]. Uncertainties represent either 2σ of the mean of measurements or 2σ of the sample analysis if the group only consists of one sample. If absent, the error bars are smaller than the symbol size.*

As a three-isotope plot cannot be used to accurately determine the cosmochemical history of N, correlations of its isotopic variations with nuclides having well-determined cosmochemical pathways can be helpful in tracking the evolution of the earliest N reservoirs. Combining mass-independent isotopic signatures of non-volatile siderophile elements (Mo, Ni, W, and Ru) with N isotope ratios, we show that CC and NC iron meteorites plot in compositionally distinct clusters in $\delta^{15}$N-$\epsilon^{64}$Ni, $\delta^{15}$N-$\epsilon^{183}$W, $\delta^{15}$N-$\epsilon^{94}$Mo, and $\delta^{15}$N-$\epsilon^{100}$Ru space (Fig. 2, Supplementary Table 2; where ε represents parts per $10^4$ deviation relative to the terrestrial standards). NC iron meteorites have $\delta^{15}$N <0 ‰ and CC iron meteorites have $\delta^{15}$N >0 ‰. This shows that the parent bodies of NC iron meteorites with a purported inner Solar System origin sourced their N from a $^{15}$N-poor environment, while those of CC iron meteorites sourced their N from a $^{15}$N-rich environment (Figs.1 and 2). This observation is in line with the heterogeneous distribution of the isotopes of non-volatile elements and provides the first evidence of CC and NC reservoirs containing isotopically distinct N during the growth of the earliest protoplanets.

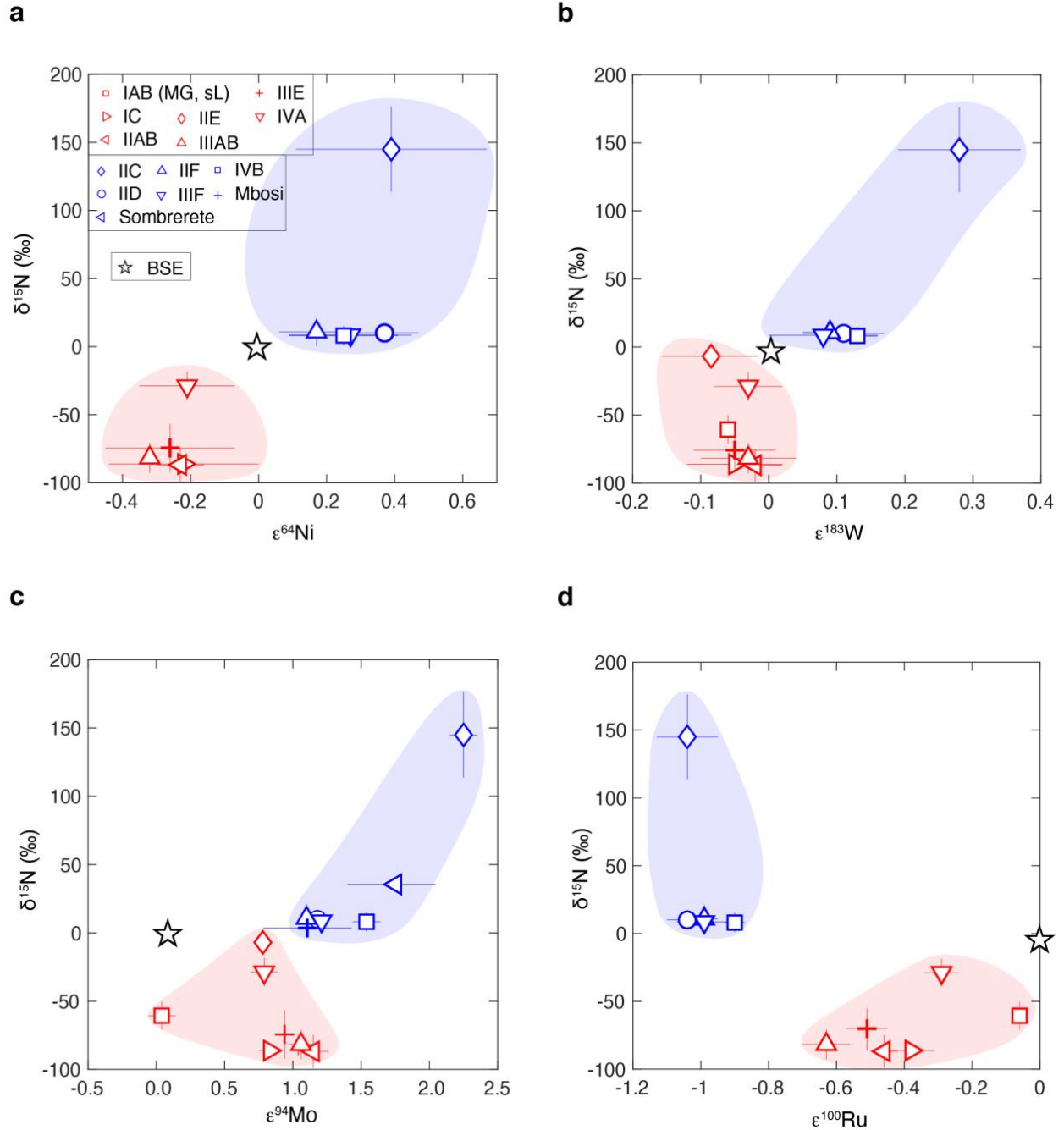

*Figure 2: CC-NC dichotomy of iron meteorites plotted in $\delta^{15}N$-$\varepsilon^{64}Ni$, $\delta^{15}N$-$\varepsilon^{183}W$, $\delta^{15}N$-$\varepsilon^{94}Mo$, and $\delta^{15}N$-$\varepsilon^{100}Ru$ space.* Carbonaceous (CC, blue) and non-carbonaceous (NC, red) iron meteorites plot in two distinct clusters. Based on their Mo isotopic signatures, iron meteorites genetically linked to NC reservoir (IAB (MG, sL), IC, IIAB, IIE, IIIAB, IIIE, and IVA) define one cluster, and others linked to CC reservoir (IIC, IID, IIF, IIIF, and IVB) along with ungrouped iron meteorites – Mbosi and Sombrerete (only in $\delta^{15}N$-$\varepsilon^{94}Mo$ space) – define the other cluster.

*The CC cluster is enriched in nuclides (Ni, W, Mo, and Ru) produced in neutron-rich stellar environments. The bulk silicate Earth (BSE) plots between CC and NC reservoirs in (a) $\delta^{15}N$-$\varepsilon^{64}Ni$ and (b) $\delta^{15}N$-$\varepsilon^{183}W$ space. On the other hand, the BSE represents an end-member composition in (c) $\delta^{15}N$-$\varepsilon^{94}Mo$ ($\varepsilon^{94}Mo$ values in CC and NC iron meteorites have variable s-process deficits relative to the terrestrial value) and (d) $\delta^{15}N$-$\varepsilon^{100}Ru$ space. Even though $^{58}Ni$ is the carrier of Ni nucleosynthetic anomalies, $\varepsilon^{64}Ni$ is used in literature to show these variations because anomalies are more precisely measured for this normalization scheme[34]. Uncertainties represent two standard deviations ($2\sigma$) from the mean of measurements (refer to Supplementary Table 2). If absent, the error bars are smaller than the symbol size.*

Does the CC-NC dichotomy of N isotopes reflect the N isotopic heterogeneity in the nebular gas, or does it reflect N being hosted in N-bearing non-volatile phases akin to other non-volatile elements? N isotopic analyses of the modern solar wind ($\delta^{15}N = -407 \pm 7$ ‰)[6], Jupiter's atmosphere ($\delta^{15}N = -375 \pm 160$ ‰)[23] and CAI hosted TiN (the first solid N-bearing phase to condense from the nebular gas; $\delta^{15}N = -364 \pm 24$ ‰)[24] all point towards an extremely $^{15}N$-poor nebular composition, the best estimate being $\delta^{15}N = -383 \pm 8$ ‰[6] (Fig. 1). Using thermodynamic models, we show that nebular ingassing (either during nebular condensation of primitive Fe, Ni-alloy before parent body accretion or equilibration of molten alloy with the nebular gas post-disruption of the parent body) can provide at most ~0.1 ppm N into the alloy, which is lower than the N content of the most N-poor iron meteorites (Fig. 3a; see Methods for details). This means that iron meteorites did not directly capture the N isotopic composition of the nebular gas. Therefore, an intermediate parent body process is required to explain N abundances in iron meteorites. N can partition into the metallic cores during core-mantle differentiation in rocky planetary bodies[2,25]. For oxygen fugacities ($fO_2$s) relevant during core-mantle differentiation in the parent bodies of iron meteorites[26] we show that N abundances in iron meteorites can be explained by partitioning of N into the core forming Fe, Ni-alloy melts owing to higher $P_N$ during core-mantle differentiation (Fig. 3a; see Methods). This holds true within the predicted size range of their parent bodies[27] along with variations in the composition of accreting material (tracked by $fO_2$ of core-mantle differentiation)[26], depending on their accretion zones in the protoplanetary disk (Fig. 3b; Extended Data Fig. 3).

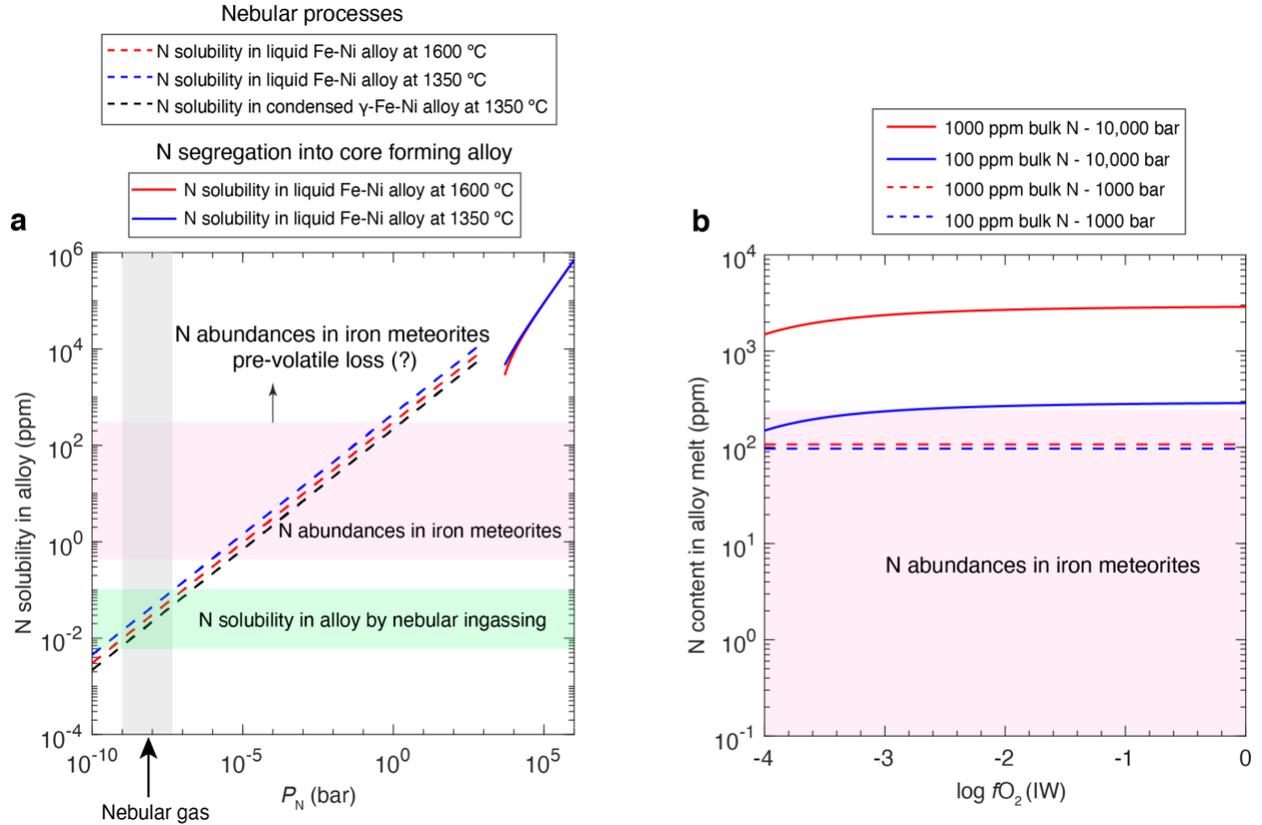

*Figure 3: N content in iron meteorites can be explained by the segregation of N into protoplanetary cores rather than via nebular ingassing into Fe, Ni-alloys. a) N solubility in solid Fe, Ni-alloys and Fe, Ni-alloy melts as a function of partial pressure of N ($P_N$). Typical $P_N$ in nebular gas is too low to ingas sufficient amounts of N into Fe, Ni-alloys and explain the N content in iron meteorites (see Methods for details). $P_N$ during core-mantle differentiation can be 13 orders of magnitude higher than the nebular pressure (due to core-mantle differentiation at the base of magma oceans for asteroid-sized bodies)[35]. Therefore, core forming Fe, Ni-alloy melts have the capacity to incorporate >10,000 ppm N depending on the pressure of alloy-silicate equilibration. b) Segregation of N into Fe, Ni-alloy melts can explain N abundances for a variety of core-mantle differentiation scenarios. Availability of modest amounts of bulk N (100 and 1000 ppm) during core-mantle differentiation can incorporate ~100 ppm of N in Fe, Ni-alloy melts at 1000 bar and ~200-3000 ppm of N in the alloy at 10,000 bar. For core-mantle differentiation at 1000 bar (for a Vesta-sized body), N solubility rather than $D_N^{alloy/silicate}$ sets the N content in the alloy melts. While at 10,000 bar, N solubility in the alloy melts is much higher (~8500 ppm; therefore, $D_N^{alloy/silicate}$ sets the N content in the alloy under those conditions (see*

*Methods for details)). Temperature is fixed at 1600 °C, alloy/silicate mass ratio at 0.5 and $fO_2$ is varied within the predicted range for alloy-silicate equilibration in the parent bodies of iron meteorites[26]. The pink shaded regions represent the range of N contents in iron meteorites.*

Because the accretion zones of NC iron meteorites lie within the thermally processed inner disk[8,28], non-volatile dust carriers or refractory organics, rather than ices or labile organics, should have been the primary source of N in their parent bodies. This is contrary to the conventional classification of N as a highly volatile element during protoplanetary disk processes. Enstatite chondrites attest for the presence of hundreds of ppm N in refractory phases like osbornite (TiN), nierite ($Si_3N_4$), sinoite ($Si_2N_2O$), and isostructural substitution for O in silicate lattice[14]. This is in agreement with the presence of noble gases (even more volatile than N) in enstatite chondrites[29], likely to be hosted by refractory phases related to organic precursors[22]. N-bearing refractory organics have also been identified in carbonaceous chondrites, comets, and interstellar medium[3]. Production of N-bearing refractory organics via photon-gas interactions within the protoplanetary disk[22,30], and/or as a heritage of the molecular cloud[21], followed by widespread dispersal via turbulent diffusion could have efficiently distributed organic precursors in the mid-plane of the disk (see Methods for details).

Irrespective of the cause of its origin, the CC-NC dichotomy for N isotopes in iron meteorites portrays a nuanced picture for the distribution of N in the early Solar System. A distinct N isotopic composition of the parent bodies of iron meteorites and CAIs[31] growing concomitantly with $^{15}$N-poor Sun and Jupiter shows that the N reservoirs from which formed the earliest protoplanets formed accreted their N were isotopically decoupled from the nebular reservoir (Fig. 1). Even though most of the N in the protoplanetary disk was hosted by nebular gas, rocky protoplanets forming on similar timescales in the inner and outer Solar System accreted their N from much smaller, non-volatile reservoirs. The enrichment of $^{15}$N in non-volatile carriers relative to nebular gas and a $^{15}$N gradient from the inner to the outer Solar System in these carriers either predated, or was synchronous, with the growth of the earliest formed protoplanets. This has important implications for the transport of N, and potentially other volatiles, in the planet forming region of the early Solar System, which needs to be further explored in future models on the dynamics of early Solar System. Distinct N isotopic signatures in CC and NC iron meteorites coupled with extremely short accretion timescales of their parent bodies (less than ~0.3 and ~0.9 Myr for NC and CC, respectively)[7] provides the first conclusive evidence that the earliest formed protoplanets in the inner Solar System sourced their N from an isotopically distinct reservoir.

An evidence for the presence of isotopically distinct N in CC and NC iron meteorites, with their separation at $\delta^{15}N = 0$ ‰, has important implications for the origin of volatiles in the bulk silicate Earth (BSE = mantle + crust + atmosphere). Using the estimated abundance[32] and isotopic composition[33] of N in the mantle ($\delta^{15}N_{mantle} = -5 \pm 4$ ‰ (2 s.d.) represents the best estimate accounting for N isotope heterogeneity in the Earth's mantle), atmosphere ($\delta^{15}N_{atmosphere} = 0$ ‰), and crust ($\delta^{15}N_{crust}$ ~6 ‰), $\delta^{15}N_{BSE}$ is estimated to be $-1.5 \pm 3$ ‰ (2 s.d.). This value lies between those of CC and NC reservoirs, suggesting that N in the BSE likely reflects a mixture of CC and NC material (Fig. 1 and 2). Using mass balance calculations with inverse Monte-Carlo simulations, we predict that the BSE's N isotopic composition can be reproduced from an admixture of CC and NC material – a) 59 % CC and 41% NC (13 %; 1 s.d.) by mass (assuming similar N abundances in the CC and NC material), or b) 30 % CC and 70 % NC (15 %; 1 s.d.) by mass (assuming N abundances in the CC and NC material in a ratio of 4:1, based on the N contents of CI and enstatite chondrites, which are assumed to be representative of the CC and NC reservoirs, respectively). Although N in the present-day BSE has a mixed CC-NC reservoir heritage, it is not possible to directly constrain the stage of Earth's growth at which the BSE acquired its present-day N (as estimated for non-volatile elements in ref. 12). This is due to the expected N loss to space during Earth's protracted growth history[25]. A moderately siderophile and highly volatile character (during planetary processes) of N suggests that the early accreted N in the proto-Earth was likely segregated to the core or lost to space[25]; thus, N composition of the present-day BSE is potentially biased towards the last stages of Earth's growth. Similar conclusions of the BSE gaining its volatile inventory during the final stages of Earth's accretion (either during the Moon forming event and/or late accretion) via an admixture of CC and NC reservoirs have also been made using its C/N ratio and Mo isotopic composition as constraints[2,28]. The origin of isotopically distinct N in the inner Solar System protoplanets may mean that C and water were also available in similar accreting zones via transport of organics and dust in the disk. However, the timing of their origin in large, Earth-like planets was potentially linked to the movements of protoplanets and planetary embryos and not to the transport of dust and organics.

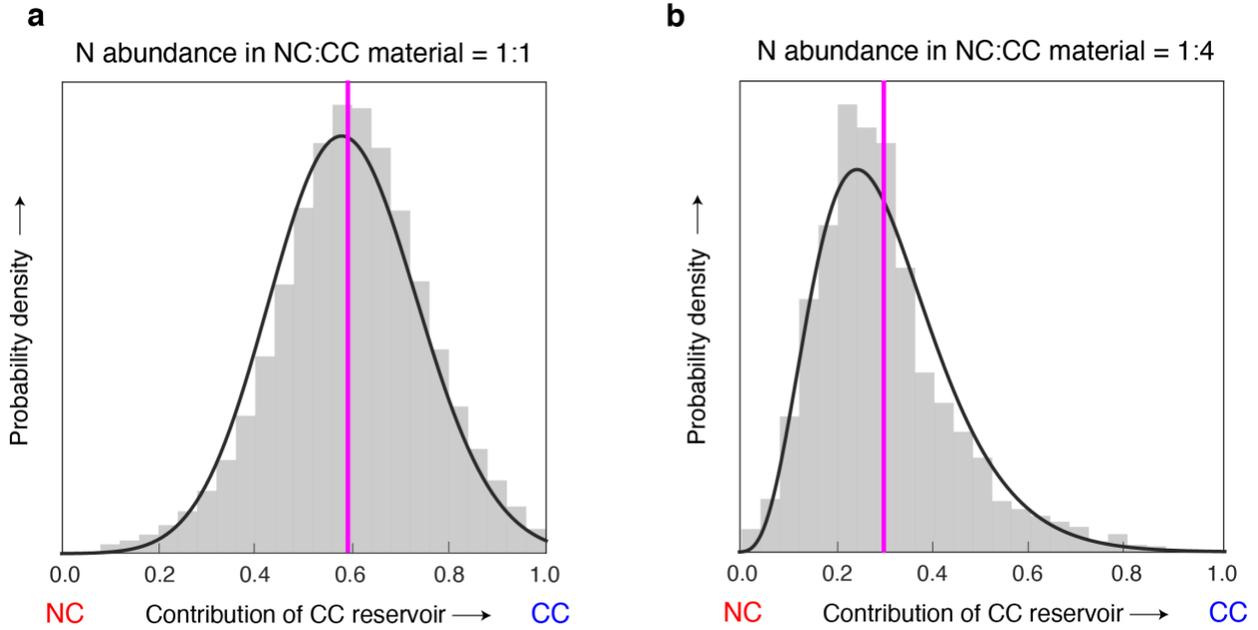

*Figure 4: Contribution of CC and NC material to the N budget of the BSE.* $10^6$ *inverse Monte Carlo simulations are used to predict the contribution of CC and NC material to the present-day N inventory of the BSE for **a**) CC and NC material having similar N abundances **b**) CC and NC material having N in the ratio of 4:1, i.e., similar to the N abundance ratio in CI to enstatite chondrite meteorites[14]. Pink lines represent the mean of the contribution of CC and NC material to the N budget of the present-day BSE.*

**Acknowledgements**

Cin-Ty Lee is acknowledged for fruitful discussions during the early stage of this research. Anurodh Vyas and Johnny D. Seales helped with the CC-NC reservoir mixing calculations for the BSE. Amrita P. Vyas helped improve the clarity of our communication. **Funding:** D.S.G. received support from NASA FINESST grant 80NSSC19K1538 and Lodieska Stockbridge Vaughn Fellowship by Rice University. R.D. was supported by NASA grants 80NSSC18K0828 and 80NSSC18K1314. BM was supported by the European Research Council grant PHOTONIS 695618.

**Author Contributions**

D.S.G. conceived the project, compiled the data and developed thermodynamic models. D.S.G., R.D., and B.M. interpreted the data. D.S.G. wrote the manuscript with inputs from R.D. and B.M.

**Additional information**

The data that support the plots within this paper and other findings of this study are available in Supplementary information which is available in the online version of the paper. Reprints and permissions information is available at www.nature.com/reprints. Correspondence and requests for materials should be addressed to D.S.G. (dsg10@rice.edu).

**Competing financial interests**

The authors declare no competing financial interests.

**Data availability**

The authors declare that the data supporting the findings of this study are available within the article and its Supplementary Information files.

# Methods

**Abundances and isotopic ratios of N in iron meteorites**

We compiled abundances and isotopic ratios of N in iron meteorites reported in literature (Supplementary Table 1, Extended Data Fig. 1)[19,20,36–45]. N abundances and isotopic ratios have been measured in both fractionally crystallized groups (formed by fractional crystallization of fully molten Fe, Ni alloys; belonging to IC, IIA, IIAB, IIC, IID, IIF, IIIAB, IIIE, IIIF, IVA, and IVB groups) and silicate bearing groups (formed from partially molten Fe, Ni alloys with incomplete separation from silicate reservoirs; belonging to IAB (MG, sL) and IIE groups) (for iron meteorite classification refer to ref. 46). N was reported to be predominantly (>98%) present in the Fe, Ni-alloy phases (kamacite and taenite) with a minor presence in accessory phases (nitrides, phosphides, carbides, and graphite). The distribution of N contents and isotopic ratios are in Extended Data Figure 1 and 2 and described in the text. Kernel smoothing function, an inbuilt distribution fit function in MATLAB®, was used to ascertain the statistical distribution of the data.

**Effects of planetary and cosmic processes on the alteration of $^{15}N/^{14}N$ ratios of iron meteorites**

To evaluate whether N isotopic ratios in iron meteorites capture primordial heterogeneities in the solar nebula, it is important to constrain the extent to which the ratios have been affected by planetary processes. Fractional crystallization is an important parent body process that can affect elemental distribution and also cause isotopic fractionation between solid and liquid phases[47]. In literature, the effect of fractional crystallization is quantified through iridium content in iron meteorites as it is extremely sensitive to the extent of fractional crystallization, with later crystallizing alloys being depleted in iridium as compared to early crystallizing alloys[46,47]. The absence of any systematic variations between N and iridium content as well as $\delta^{15}N$ and iridium content has been shown to imply that there is minimal elemental and isotopic fractionation, respectively, between the solid and liquid alloys during fractional crystallization[19,20].

Diffusive loss of N from the molten alloy during a degassing episode with its associated kinetic isotope fractionation can also affect N abundances and isotopic ratios in iron meteorites. Although N abundances in iron meteorites in any given group are variable, there are no positive correlations between N concentrations and $\delta^{15}$N within any given group (Extended Data Fig. 2). Therefore, within any given group, loss of N owing to degassing likely proceeded without any significant mass-dependent N isotope fractionation[19,20,44]. Also, no known kinetic or equilibrium isotope fractionation process can explain the large inter-group N isotopic variation in iron meteorites[19,20]. Although N isotopes can be fractionated to extreme limits at extremely low temperatures (<100 K), the typical temperatures associated with the loss of N in iron meteorites (possibly due to volatile loss in disrupted cores) are much higher[19,20].

Recent experimental studies that have tried to constrain alloy-silicate N isotope fractionation at 1400-1800 °C have yielded conflicting results. These studies have argued for both little (~1-5 ‰; ref. 48) and large (~25-257 ‰; ref. 49) isotope fractionation. The limited experimental data till date[48,49] suffer from large amount of N loss during the experiments (a similar loss has been observed in all previous studies[2,25,35,50] that have measured elemental fractionation of N between alloys and silicates using experimental setups similar to the above mentioned studies[48,49]), which means additional factors like N isotope fractionation due to volatile loss may have severely compromised the results[49]. Also, all theoretical calculations predict limited isotope fractionation during core-mantle differentiation[51–53]; therefore, more experimental evidence is required to reliably constrain whether large N isotopic fractionation is possible during core-mantle differentiation. Aligning with the conventional views backed by several studies[51–53], in this study we assume limited N isotope fractionation during alloy-silicate differentiation[48] such that N isotopic composition of the iron meteorites can be directly traced to the N isotopic composition of the bulk material that made up their parent bodies. If there was significant fractionation of N isotopes during alloy-silicate equilibration[49], then the material accreted by the parent bodies of iron meteorites should have had higher $^{15}$N/$^{14}$N ratios relative to those measured in the iron meteorites themselves. However, as long as the alloy-silicate differentiation for all magmatic iron meteorites occurred at largely similar $f$O$_2$ conditions[26,54], the only estimate that would be impacted are the contribution of CC and NC material to the N budget in the BSE. However, the main finding of this study that the CC and NC reservoirs had isotopically distinct N would still hold true.

Other secondary processes like solar wind implantation and cosmic ray exposure also cannot explain inter-group variation of N isotope ratios in iron meteorites[19,20]. The effect of cosmic ray spallation on the N isotopic composition of the silicates of stony meteorites with large exposure ages can be quite pronounced because cosmic rays produce $^{15}N/^{14}N$ by spallation on O (ref. 55). Because iron meteorites contain negligible O, the effect of cosmic ray exposure on N isotopic composition of iron meteorites is minimal despite their large exposure ages[19].

Consequently, a minimal effect of post-accretional processes on N isotopic ratios in iron meteorites may suggest that their N isotopic variations reflect either nebular heterogeneities of nucleosynthetic origin[19], or N isotope fractionation in the solar nebula[22,30,56,57], and these heterogeneities may have survived large scale homogenization during thermal processing in the protoplanetary disk.

**How did iron meteorites attain their N?**

A better understanding of the processes that set N abundances in iron meteorites is critical to track the isotopic evolution of N bearing reservoirs in the protoplanetary disk. If the iron meteorites acquired their N budgets before disruption of their parent bodies during core-mantle differentiation, then N isotopic ratios in the iron meteorites would inform us about the isotopic composition of the non-volatile carriers, which could be an admixture of Solar System condensed matter and pre-solar carriers. On the other hand, if N originated in the iron meteorites either prior to accretion of their parent bodies (during condensation of the Fe, Ni-alloy from a cooling nebular gas via equilibration of condensed Fe, Ni-alloy with the nebular gas) or after disruption of their parent bodies (via equilibration of the molten Fe, Ni-alloy with the nebular gas), then isotopic ratios of N in the iron meteorites would capture the N isotopic composition of nebular gas. To differentiate between these two widely different mechanisms of possible acquisition of N by iron meteorites, we use thermodynamic calculations to assess the likelihood of each process.

*Incorporation of N into Fe, Ni-alloy via nebular gas interaction*

Here we first test whether equilibration of Fe, Ni-alloy with nebular gas can set N abundances in iron meteorites. In the metallurgy literature, N dissolution in metals equilibrating with molecular N has been explained using Sievert's law. Applying a similar framework, N solubility in the metal, either γ-iron alloy (stable form of iron at temperatures relevant for condensation of Fe, Ni-alloy from the nebular gas) or molten metal, equilibrating with nebular gas can be written as:

½ $N_{2(g)}$ = $N_{(metal)}$   (Eq. 1)

$K_{eq} = \dfrac{a_N^{metal}}{f_{N_2}^{0.5}}$   (Eq. 2)

where, $K_{eq}$ is the equilibrium constant for the reaction, $a_N^{metal}$ is the activity of N in the metal and $f_{N_2}$ is the fugacity of $N_2$ in the gaseous phase. Although $N_2$, $NH_3$, and HCN are the potential N carriers in nebular gas, the proportion of N present as $N_2$ would be at least comparable, if not greater than, to $NH_3$ and HCN at <10 AU in the protoplanetary disks[58,59]. Therefore, meaningful information can be gathered by tracking incorporation of N into the metals via equilibration with $N_2$ only in nebular gas.

For an infinitely dilute solution of N in the liquid metal, Eq. 2 can be written as:

$$[N] = K_{eq} \cdot P_{N_2}^{0.5} \quad (Eq. 3)$$

where, [N] is the concentration of N in the metal and $P_{N_2}$ is the partial pressure of $N_2$ in the gaseous phase. Therefore, N concentration in the metal can be determined if $K_{eq}$, and $P_{N_2}$ in the equilibrating gas phase are known.

Nebular gas pressure in the rocky body forming region of the protoplanetary disk varied between $10^{-3}$ and $10^{-6}$ bar on time scales of ~0-2 Myr[60,61]. The nebular gas was primarily composed of molecular hydrogen ($H_2$). N abundance in the Solar System is ~4 orders of magnitude lower than H. To take into account the variations of nebular gas pressure as a function of radial distance, we vary $P_{N_2}$ between $10^{-7}$ and $10^{-10}$ bar in our calculations for the condensation of Fe, Ni alloy from the cooling nebular gas in both inner and outer Solar System. As $K_{eq}$ is not dependent on $P$, we use $K_{eq}$ values of Sieverts law determined for γ-Fe, Ni alloys[62] and Fe, Ni alloy melts[63,64] at $P_{N_2}$ in the atmospheric pressure range.

Figure 3a shows N solubility in γ-Fe, Ni solid alloys and Fe, Ni-alloy melts during equilibration with nebular gas. The predicted N solubility for both condensed alloys as well as alloy melts lies between 0.005 and 0.1 ppm. N solubility range predicted by our calculations is an upper bound because it assumes that the entire parcel of alloy equilibrated with the nebular gas and there was no kinetic barrier for N diffusion into the metallic phase. Although these assumptions might hold true for equilibration of nebular gas with the condensing Fe, Ni-alloy, it might not be realistic for post-disruption equilibration of stripped metallic cores with nebular gas owing to their relatively large sizes as well as rapid crystallization rates of molten alloy post-crystallization[65]. However, even this upper bound is one to three orders of magnitude lower than the present-day N abundances in iron meteorites (Fig. 3). Therefore, equilibration of Fe, Ni-alloy with nebular gas, either pre-accretion of parent bodies of iron meteorites during condensation of molten alloy from nebular gas, or post-disruption of parent bodies of iron meteorites in the presence of nebular gas, cannot set the N abundances in iron meteorites. A similar conclusion was also reached by ref. 66 which used thermodynamic equilibrium calculations for N exchange between metal and nebular gas.

*Incorporation of N into Fe, Ni-alloy during core-mantle differentiation*

To test whether segregation of N into the core forming alloy may explain the observed N abundances in iron meteorites, the knowledge of thermodynamic parameters, i.e., *P*, *T*, and *f*$O_2$, applicable for differentiation of the parent bodies of iron meteorites is required. The sizes of parent bodies of various groups of iron meteorites as well as the composition and depth of their alloy-silicate equilibration are not well constrained. Using cooling rates for several groups of iron meteorites, it has been assumed that iron meteorites were formed by isobaric cooling of metallic cores inside differentiated parent bodies, which had diameters between 20 and 350 kms (see ref. 67 and references therein). However, a recent revision of metamorphic cooling rates of IVA iron meteorites was used to predict that the metallic cores of the parent bodies of several magmatic and non-magmatic iron meteorites could have had diameters of ~300 km[27,68] and these cores would have been exposed post hit and run collision between Moon- to Mars-sized protoplanets[69]. Therefore, several groups of iron meteorites can be the remnants of cores of differentiated protoplanets that had diameters of ~500-1000 km or larger[27].

4 Vesta, the most well studied asteroid in our Solar System, which has a mean diameter of ~525 km can be used as a proxy to understand core-mantle differentiation in the parent bodies of iron meteorites. Several constraints on the pressure-temperature-oxygen fugacity (*P*-*T*-*f*$O_2$) of its core-mantle differentiation exist in the literature utilizing geochemistry of HED meteorites. *P* of its alloy-silicate equilibration could have been close to 1000 bar (corresponding to the estimated pressure at the core-mantle boundary of 4 Vesta), *T* between 1725 and 1850 K, and mean *f*$O_2$ to be approximately 2 log units below the iron-wüstite (IW) buffer (IW–2)[70]. Similar *f*$O_2$ conditions (~IW–4 to –2) for core-mantle differentiation in the parent bodies of IAB (MG, sL) and IIE iron meteorites have also been estimated[26]. Using these estimates of *P*-*T*-*f*$O_2$ as guides, we simulate core-mantle differentiation in parent bodies of iron meteorites at *P* = 1000 and 10000 bars, *T* = 1873 K, and log*f*$O_2$ = IW–4 and IW–2. To estimate the amount of N that can be dissolved into the core forming alloy during alloy-silicate equilibration, we use the parametrized $D_N^{alloy/silicate}$ relationship from ref. 25. The bulk N content available in the parent bodies during alloy-silicate equilibration was fixed at 100 ppm and 1000 ppm (based on the variation of N content in enstatite and carbonaceous chondrites[14]), while the alloy/silicate mass ratio of the parent bodies was varied between 0.01 and 1. The entire parcel of the core forming alloy is assumed to undergo 100% equilibration with the molten silicate of the parent bodies because alloy melt-silicate melt equilibration has been shown to be extremely efficient in small rocky bodies[71]. Finally, if the concentration of N set by alloy-silicate equilibration is greater than the solubility limit of N in the core forming alloy set by Sievert's law formulation (using parameterized relationship for N solubility in the alloy melts determined by ref. 64 at 1000 bar and ref. 35 at 10000 bar), then the final N abundance in the cores of the parent bodies would be set by the solubility limit of N in the core forming alloy rather than by $D_N^{alloy/silicate}$.

Extended Data Fig. 3 shows the measured N abundances in the core forming alloy melts as a function of the core/mantle mass ratio. N shows siderophile character at IW–2 ($D_N^{alloy/silicate}$ = ~20) and lithophile character at IW–4 ($D_N^{alloy/silicate}$ = ~0.1)[25,50]. For alloy-silicate equilibration at 1000 bar, N content in the core forming alloy at IW–2 is fixed by N solubility limit at 97 ppm for 100 and 1000 ppm bulk N, while at IW–4, N content in the alloy varies between 10 and 18 ppm for 100 ppm bulk N and it is fixed by N solubility in the alloy at 97 ppm for 1000 ppm bulk N. During alloy-silicate equilibration at 10,000 bar, N content in the alloy at IW–2 varies between ~190 and 8500 ppm for 100 and 1000 ppm bulk N, while at IW–4, N content in the alloy varies between ~10 and 180 ppm. These abundances of N in the core forming alloy lie well within the range or higher than the measured N abundances in the iron meteorites for a wide range of thermodynamic parameters. Even if the present-day N abundances in the iron meteorites represent a lower limit, with some amount N being lost during de-volatilization post disruption of their parent bodies, our calculations can yield N contents at or above the measured values depending upon the minor adjustment of thermodynamic parameters. Therefore, segregation of N into the cores of parent bodies can set the N contents in the iron meteorites for parent bodies of varying size that could have accreted different budgets of N (within the range of N content in enstatite and carbonaceous chondrites[14]) for widely different compositions of accreting material (tracked by $fO_2$) due to the variations in their accreting zones in the protoplanetary disk.

Even though the explanation of N contents in iron meteorites via core-mantle differentiation is directly applicable for magmatic irons, it is possible that a similar mechanism may be applicable for non-magmatic irons as well. The origin of non-magmatic irons is currently debated with their origins in partially differentiated bodies and/or bodies partially melted by impact[72,73]. However, the term "non-magmatic" may be misleading because it has been argued that the metals of non-magmatic irons were also once molten but could not efficiently segregate from the silicate melts[46]. If this holds true, then similar to the parent bodies of magmatic iron meteorites alloy-silicate equilibration should have also taken place in the parent bodies of non-magmatic iron meteorites, albeit without efficient alloy-silicate separation. In such a scenario, alloy-silicate equilibration could have set the N abundances in the metallic alloys of both magmatic and non-magmatic iron meteorites. However, if melting was impact induced, then the conditions of metal-silicate equilibration may deviate from those considered here. But given that $P$ has a weak effect on $D_N^{alloy/silicate}$ [25], we do not think that the conclusions drawn from the results of Figure 3b and Extended Data Fig. 3 would be affected significantly.

**Origin of N isotope heterogeneity – inherited from the molecular cloud or self-shielding processes in the protoplanetary disk?**

Did the molecular cloud start with a uniform $^{15}N/^{14}N$ ratio and later processes in the protoplanetary disk lead to the observed heterogeneities? Or do these heterogeneities have a nucleosynthetic origin with the gaseous and dust components of the parent molecular cloud having different $^{15}N/^{14}N$ ratios? $^{14}N$ and $^{15}N$ are produced by completely different pathways during stellar nucleosynthesis ($^{14}N$ by cold CNO cycle in low to intermediate mass stars/hot CNO cycle in massive stars, and $^{15}N$ during hot CNO cycle in novae or neutrino spallation on $^{16}O$ in type II supernovae[56]). The parent molecular cloud, similar to non-volatile elements[18,34], may host a heterogenous distribution of N isotopes. The clustering of $^{15}N$-rich CC iron meteorites with nuclides of W, Mo, Ru, and Ni produced in neutron-rich environments (Fig. 2)[8,18,28,34] indicates that a nucleosynthetic origin of N isotopic variation resulting from the heterogeneities of N isotopes in the parent molecular cloud is a plausible explanation. A similar explanation has been posited for the variation of K (a moderately volatile element) isotopes in meteorites[74]. However, N isotopic compositions of iron meteorites do not show any intra-cluster correlations (Fig. 2), while similar correlations are observed for non-volatile elements (for example, in Mo isotopes expressed as s-process variations) in both or at least one of the reservoirs[75]. Furthermore, N isotopic variations between CC and NC iron meteorites may vary up to 260 ‰ while the isotopes of W, Mo, Ru, and Ni vary in the order of a few epsilons, i.e., tenths of permil. The isotopic variation of non-volatile elements in pre-solar grains is in the order of hundreds of permil (e.g., Mo isotopes is SiC grains[76]), while N isotopes in pre-solar grains (e.g., SiC)[77] is in the order of thousands of permil. Thus, an exclusively nucleosynthetic origin of the N isotopic dichotomy in CC and NC reservoirs should also cause a large isotopic variation of non-volatile elements, which is not observed in the meteorite record. Also, if the variation of N isotopes was caused exclusively via nucleosynthetic anomalies, a comparable variation of C isotopes in meteorites should be expected because N and C both have similarly large isotopic variation in presolar grains (e.g., SiC)[77]. However, this is also not observed in the meteorite record where $\delta^{13}C$ varies at less than 40 ‰[14]. In combination, these reasons make a nucleosynthetic origin of N isotopic dichotomy less likely.

Observations of large N isotope variations in organic matter[3] have been linked to N isotope fractionation in cold, dense molecular cloud cores[78]. Alternatively, the synthesis of refractory organic matter in the ionized gas phase of the solar nebula has been posited as an explanation for the origin of N isotope heterogeneities in the protoplanetary disk[22,30]. In contrast to a cold-environment origin in the interstellar medium[79], photodissociation of $N_2$ in the nebular gas by interstellar or stellar UV-X rays anywhere at the surface of the disk[30] (including its warm regions)[22] followed by recombination with surrounding H and C atoms could have led to the formation of $^{15}N$-enriched organics and ices through self-shielding process[22,30]. The organic dust may have then settled in the dead zones of the mid-plane of the inner as well as outer regions of the disk, possibly within few tens of thousands of years[22,31,57]. The process that presumably caused $^{15}N$ enrichment of the disk were more pronounced in the CC reservoir that was further out in the colder part of the disk. Therefore, either these processes were more efficient there, or $^{15}N$-rich carriers were more likely to survive those regions.

**Mixing calculations for N isotopic composition of the BSE**

We computed the mass fraction of NC and CC materials necessary to reproduce the N isotopic composition of the BSE. The mixing equation is:

$$[(m_N)_{NC} + (m_N)_{CC}] \times (\tfrac{^{15}N}{^{14}N})_{BSE} = (m_N)_{NC} \times (\tfrac{^{15}N}{^{14}N})_{NC} + (m_N)_{CC} \times (\tfrac{^{15}N}{^{14}N})_{CC} \quad \text{(Eq. 4)}$$

where $(m_N)_{NC}$ and $(m_N)_{CC}$ is the mass of N inherited from NC and CC materials, respectively. $(\tfrac{^{15}N}{^{14}N})_{BSE}$, $(\tfrac{^{15}N}{^{14}N})_{NC}$, and $(\tfrac{^{15}N}{^{14}N})_{CC}$ are the N isotopic ratio in the BSE, NC, and CC reservoirs, respectively. $(\tfrac{^{15}N}{^{14}N})_{BSE} = (3.675 \pm 0.001) \times 10^{-3}$ (details in main text), $(\tfrac{^{15}N}{^{14}N})_{NC} = (3.654 \pm 0.012) \times 10^{-3}$ (unweighted mean of the average values of NC iron meteorite groups in Supplementary

Table 2), and $(\frac{^{15}N}{^{14}N})_{CC} = (3.688 \pm 0.011) \times 10^{-3}$ (unweighted mean of the average values of CC iron meteorite groups as well as ungrouped CC irons in Supplementary Table 2). Errors represent 1 sigma deviation from the mean.

By rearranging, Eq. 4 can be written as:

$$(\frac{^{15}N}{^{14}N})_{BSE} = \frac{(m_N)_{NC}}{[(m_N)_{NC} + (m_N)_{CC}]} \times (\frac{^{15}N}{^{14}N})_{NC} + \frac{(m_N)_{CC}}{[(m_N)_{NC} + (m_N)_{CC}]} \times (\frac{^{15}N}{^{14}N})_{CC} \quad \text{(Eq. 5)}$$

$$(\frac{^{15}N}{^{14}N})_{BSE} = \frac{(M)_{NC} \times [N]_{NC}}{[(M)_{NC} \times [N]_{NC} + (M)_{CC} \times [N]_{CC}]} \times (\frac{^{15}N}{^{14}N})_{NC} + \frac{(M)_{CC} \times [N]_{CC}}{[(M)_{NC} \times [N]_{NC} + (M)_{CC} \times [N]_{CC}]} \times (\frac{^{15}N}{^{14}N})_{CC}$$

(Eq. 6)

where $[N]_{NC}$ and $[N]_{CC}$ represent N concentrations in NC and CC reservoirs, respectively. $M_{NC}$ and $M_{CC}$ represent the net mass of NC and CC reservoirs contributing to the N budget of the BSE.

Using $10^6$ inverse Monte Carlo simulations we calculated the mass fractions of NC and CC materials required to satisfy the N isotope ratios of the present-day BSE assuming: 1) $[N]_{NC} = [N]_{CC}$, i.e., equal concentration of N in NC and CC materials (Fig. 4a). 2) $[N]_{NC} = 0.25 \times [N]_{CC}$, i.e., NC and CC materials having N in the ratio of 1:4 (similar to N abundances in enstatite and CI chondrites, respectively)[80,81] (Fig. 4b).

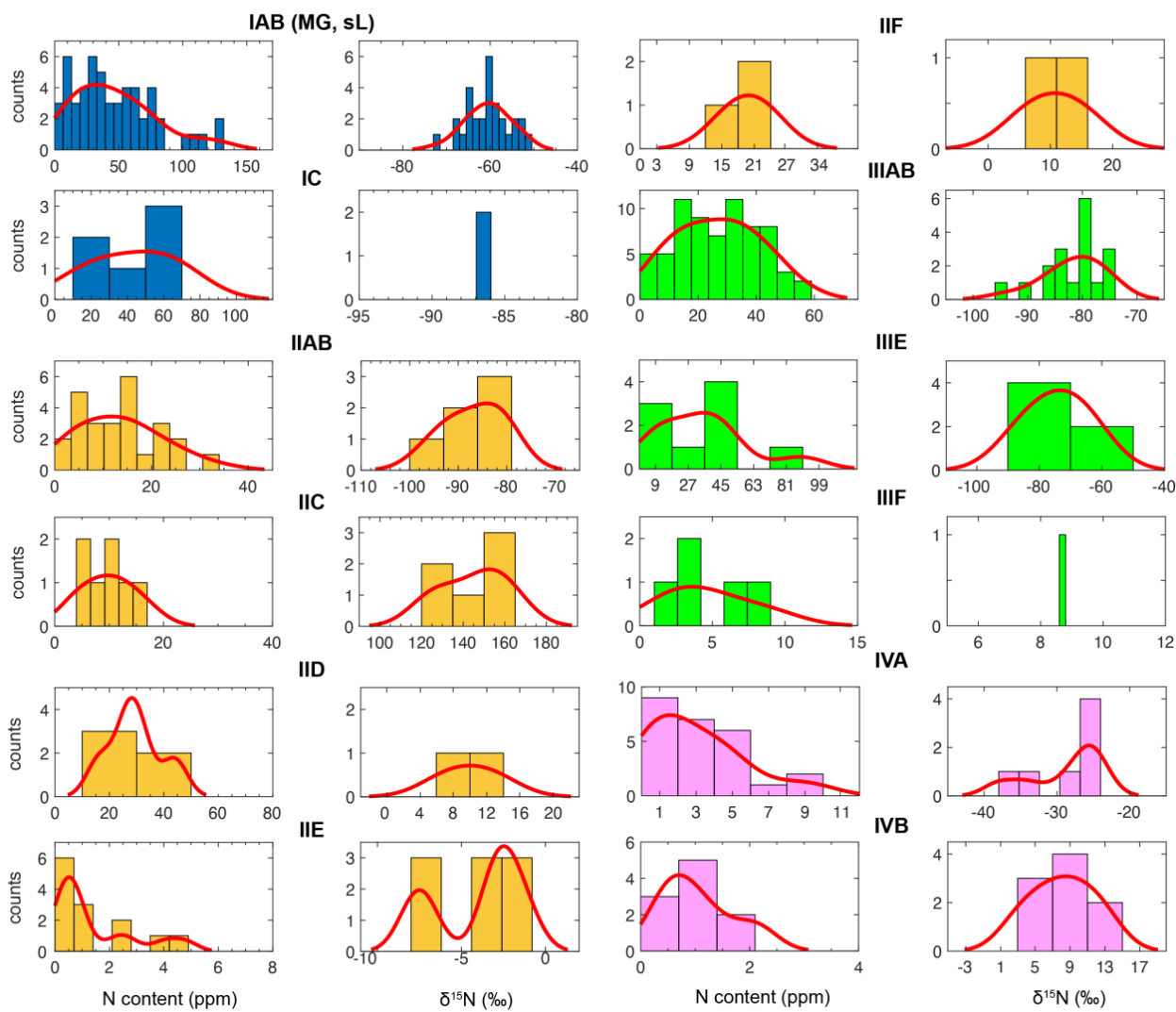

**Extended Data Figure 1: Statistical variation of N abundances and isotope ratios in different groups of iron meteorites.** Probability distributions function using kernel density distribution function (using Matlab®) are used to ascertain the statistical variation of N abundances and isotope ratios for each group of iron meteorites. For each group, $\delta^{15}N$ varies within a small range and shows sharp peaks for mean values, while N abundances show large variation.

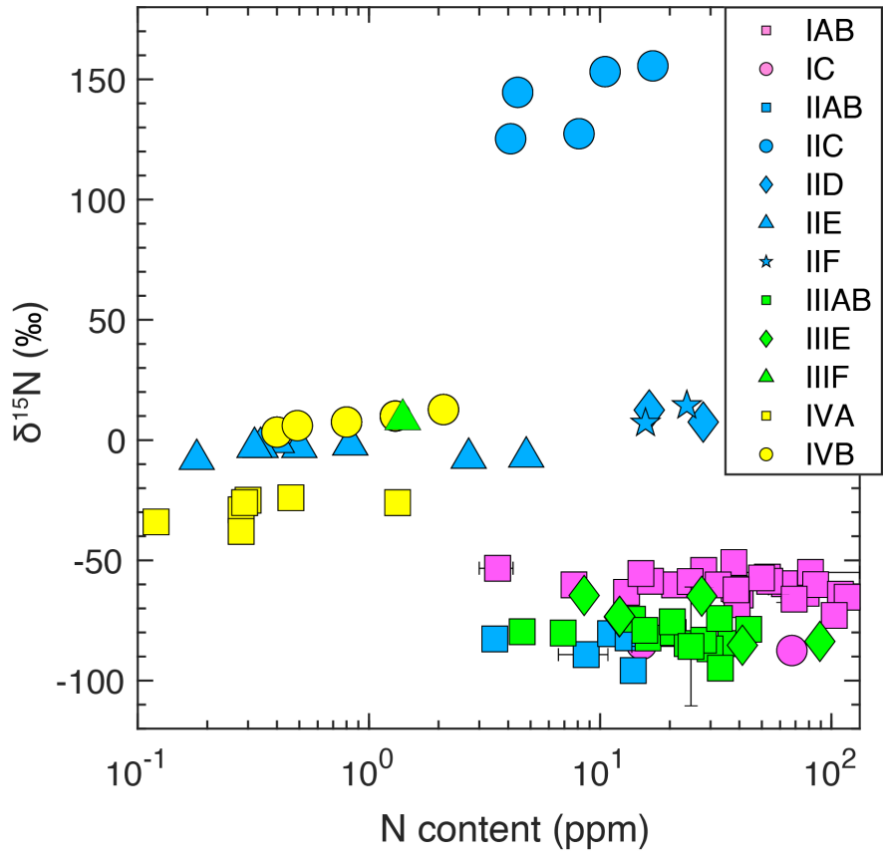

**Extended Data Figure 2: δ¹⁵N values of iron meteorites plotted against their N abundances.** For each iron meteorite group, $\delta^{15}N$ falls in a narrow range, while N contents may vary over two orders of magnitude. Lack of any relationship between $\delta^{15}N$ values and N abundances within any given group of iron meteorites argues against significant mass-dependent fractionation of N isotopes via volatility-related losses during planetary processing[19,20,43]. $\delta^{15}N$ values for a given group of iron meteorites represents the mean of the measurements calculated from the data in Supplementary Table 1.

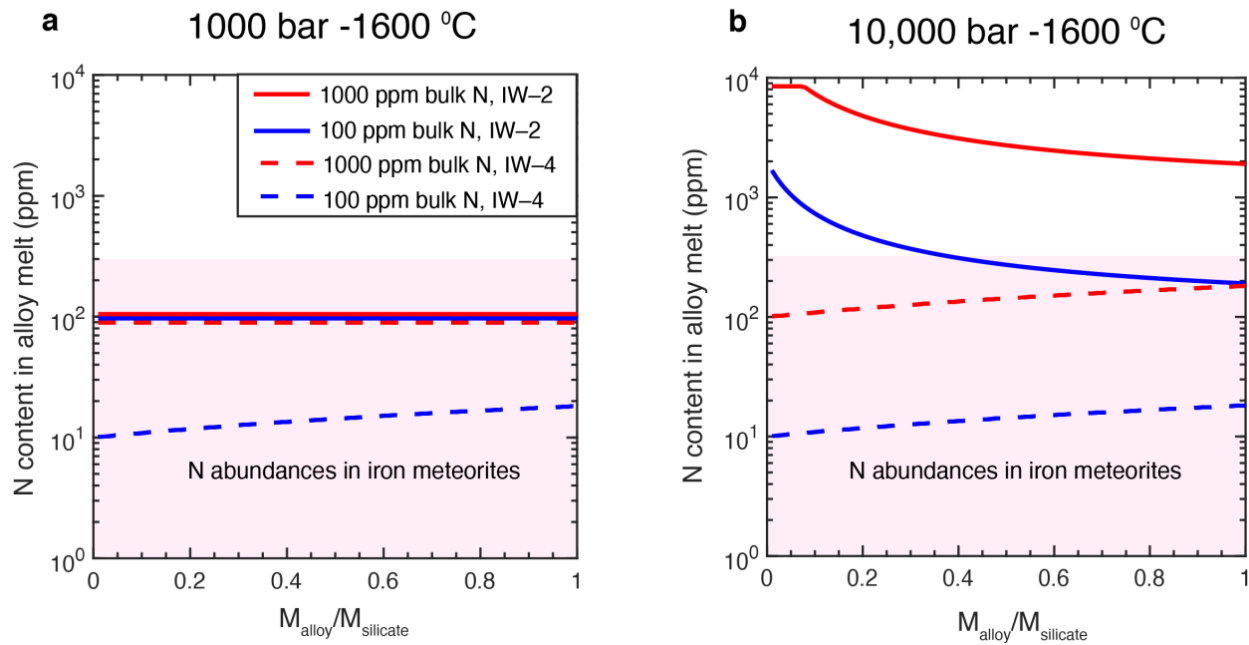

**Extended Data Figure 3: Partitioning of N into Fe, Ni-alloy melts can explain N abundances for a variety of core-mantle differentiation scenarios**. **a)** For alloy-silicate equilibration at 1000 bar-1600 °C and varying alloy/silicate mass ratio between 0 and 1 (within the range of all differentiated rocky bodies in the inner Solar System except Mercury), N content in the core forming Fe, Ni-alloy melts varies between ~10 and 100 ppm. Fixed values of N content in the alloy at ~100 ppm are owing to low N solubility in the alloy at 1000 bar. **b)** For alloy-silicate equilibration at 10,000 bar-1600 °C and varying alloy/silicate mass ratio, N content in the core forming Fe, Ni-alloy melts varies between ~10 and 10,000 ppm. The pink shaded regions represent the range of N contents in iron meteorites.